\documentclass[graybox]{svmult}

\usepackage{color}
 
\usepackage{url}

\usepackage{mathptmx}       
\usepackage{helvet}         
\usepackage{courier}        

\usepackage{type1cm}        
\usepackage{makeidx}         
\usepackage{graphicx}        
\usepackage{multicol}        
\usepackage[bottom]{footmisc}

\makeindex             

\usepackage{amsmath}
\usepackage{amssymb}
\renewcommand{\d}{{\rm d}}
\newcommand{\e}{{\rm e}}
\newcommand{\PD}[2]{\frac{\partial #1}{\partial #2}}
\newcommand{\FD}[2]{\frac{\d #1}{\d #2}}

\DeclareSymbolFont{AMSb}{U}{msb}{m}{n}
\DeclareMathSymbol{\NSet}{\mathalpha}{AMSb}{"4E}
\DeclareMathSymbol{\ZSet}{\mathalpha}{AMSb}{"5A}
\DeclareMathSymbol{\RSet}{\mathalpha}{AMSb}{"52}
\DeclareMathSymbol{\CSet}{\mathalpha}{AMSb}{"43}

\begin{document}
\title*{Next generation neural mass models}

\author{Stephen Coombes and \'Aine Byrne}
\institute{Centre for Mathematical Medicine and Biology, School of Mathematical Sciences, University of Nottingham,
University Park, Nottingham, NG7 2RD, UK.
}
%
%
\maketitle

\abstract{
Neural mass models have been actively used since the 1970s to model the coarse grained activity of large populations of neurons and synapses.  They have proven especially useful in understanding brain rhythms.  However, although motivated by neurobiological considerations they are phenomenological in nature, and cannot hope to recreate some of the rich repertoire of responses seen in real neuronal tissue.  In this chapter we consider the $\theta$-neuron model that has recently been shown to admit to an exact mean-field description for instantaneous pulsatile interactions.  We show that the inclusion of a more realistic synapse model leads to a mean-field model that has many of the features of a neural mass model coupled to a further dynamical equation that describes the evolution of network synchrony.
A bifurcation analysis is used to uncover the primary mechanism for generating oscillations at the single and two population level.  Numerical simulations also show that the phenomena of event related synchronisation and desynchronisation are easily realised.  Importantly unlike its phenomenological counterpart this \textit{next generation neural mass model} is an exact macroscopic description of an underlying microscopic spiking neurodynamics, and is a natural candidate for use in future large scale human brain simulations.}\footnote{
Contribution to the Workshop ``Nonlinear Dynamics in Computational Neuroscience: from Physics and Biology to ICT" held in Turin (Italy) in September 2015.}

\section{Introduction}
\label{sec:Introduction}

The term \textit{neural mass model} is often used to refer to low dimensional models that aim to describe the coarse grained activity of large populations of neurons and synapses.  
They are typically cast as systems of ordinary differential equations (ODEs) and in their modern incarnations are exemplified by variants of the two dimensional Wilson-Cowan model \cite{Wilson72}.  This model tracks the activity of an excitatory population of neurons coupled to an inhibitory population.  With the augmentation of such models by more realistic forms of synaptic and network interaction they have proved especially successful in providing fits to neuroimaging data.  Historically one of the first examples in this area is the Zetterberg model \cite{Zetterberg1978} for the electroencephalogram (EEG) rhythm.  This is based on previous ideas developed by Lopes da Silva and colleagues \cite{daSilva1974,daSilva1976}
and is built from three interacting neural mass models, as a minimal model of a cortical column.  The first represents a population of pyramidal cells, the second a population of excitatory interneurons, and the third a population of inhibitory interneurons.  Since its introduction the Zetterberg model has become more widely popularised by the work of Jansen and Rit \cite{Jansen95} and used to explain epileptic brain dynamics, particularly by Wendling and colleagues, as recently reviewed in \cite{Wendling2015}.  Another well known neural mass model is that of Liley, which pays particular attention to the role of synaptic reversal potentials, and see \cite{Dafilis09} for a discussion of this model within the context of Freeman's ideas on the importance of chaos for cognition and perception \cite{Freeman1992}.  
As well as proving useful for understanding EEG rhythms ranging from delta ($1-4$ Hz) through to gamma ($30-70$ Hz) \cite{Sotero2007},
neural mass models have been used to describe brain resonance phenomena \cite{Spiegler2011}, resting brain state activity \cite{Deco11} and are very popular in the neuroimaging community.  In this latter instance they are often used for model driven fusion of multiple neuroimaging modalities, such as EEG and functional magnetic resonance imaging (fMRI) \cite{Valdes-Sosa2009}, as well as to augment the dynamic causal modelling framework for understanding how event-related responses result from the dynamics of coupled neural populations \cite{Moran2013}.  Moreover, they are now an integral part of the Virtual Brain project that aims to deliver the first open simulation of the human brain based on individual large-scale connectivity \cite{Sanz-Leon2015}, as well as play a key role in the neuro-computational modelling of neurological and psychiatric disorders \cite{Bhattacharya2015}.  This latter work is especially viable since neural mass models can incorporate accurate descriptions of synaptic processing, typically in the form of a synaptic response function that is driven by firing \textit{rate} rather than by the arrival times of individual action potentials.
However, it is important to remember that at heart all neural mass models to date are essentially phenomenological, with state variables that track coarse grained measures of the average membrane potential, population firing rate or synaptic activity.  At best they are expected to provide appropriate levels of description for many thousands of near identical interconnected neurons with a preference to operate in \textit{synchrony}.  This latter assumption is especially important for the generation of a sufficiently strong physiological signal that can be detected non-invasively.  The variation of synchrony within a neuronal population is believed to underly the decrease or increase of power seen in given EEG frequency bands.  The former 
phenomenon is called event-related desynchronisation (ERD), and the latter event-related synchronisation (ERS) \cite{Pfurtscheller1999}.  Unfortunately the assumption of synchrony within neural mass models means that they cannot hope to describe ERD and ERS, at least not at the single population level.  Rather this sets a natural challenge for the next generation of neural mass models.  It is precisely this issue that we take up in this chapter.

As a starting point to move beyond the current neural mass models we draw inspiration from the physics of self-organised networks.  In particular the observation of macroscopic coherent states in  
large networks of coupled spiking neuron models has inspired a search for equivalent low-dimensional dynamical descriptions, and see \cite{Ashwin2016} for a recent review of oscillatory network dynamics in neuroscience.  However, although the mathematical step from microscopic to macroscopic dynamics has proved elusive for the majority of spiking models the $\theta$-neuron model has proven amenable to such a reduction for pulsatile coupling by Luke \textit{et al}. \cite{Luke2013}.  A similar approach by Montbri\'o \textit{et al}. \cite{Montbrio2015} has been used to reduce networks of quadratic integrate-and-fire neurons.  Here we show how to naturally augment these approaches to incorporate the biologically realistic forms of synaptic coupling that are commonly adopted within current neural mass models.  In this way we arrive at the first instance of a next generation neural mass model, with a derived (as opposed to postulated) population firing rate that couples directly to a dynamical variable describing population synchrony.  In \S \ref{sec:neuralmass} we discuss 
the main elements of synaptic processing that are incorporated within standard neural mass models, and give a heuristic description of how to close the equations of motion in terms of population firing rates.  The same model of a synapse is used in \S \ref{sec:thetanetwork}, though this time driven by the spike times arising in a network of $\theta$-neurons.  For a large globally coupled network an exact mean field description is derived, and the form of the equations compared and contrasted with standard phenomenological neural mass models.  
In \S \ref{sec:nextgeneration} we present a bifurcation analysis for the single and two population mean-field models, and use this to highlight the primary mechanisms for generating population oscillations.  Importantly we also show, through direct numerical simulations, that the model supports ERD and ERS. Finally in \S \ref{sec:discussion} we reflect upon the use of such models in future large scale human brain simulations, as well as their subsequent mathematical analysis.

\section{Neural mass modelling}
\label{sec:neuralmass}

Neural mass models generate brain rhythms using the notion of population firing rates, aiming to side-step the need for large scale simulations of more realistic networks of spiking neurons.  However, both approaches often make use of the same level of description for synaptic processing, in a manner that we shall now clarify.

At a synapse, presynaptic firing results in the release of neurotransmitters that causes a change in the membrane conductance of the post-synaptic neuron.  This post-synaptic current may be written $I = g (v_{\text{syn}}-v)$, 
where $v$ is the voltage of the post-synaptic neuron, $v_{\text{syn}}$ is its membrane reversal potential and $g$ is a conductance.  This conductance is proportional to the probability that a synaptic receptor channel is in an open conducting state.  This probability depends on the presence and concentration  of neurotransmitter released by the presynaptic neuron. The sign of $v_{\text{syn}}$ relative to the resting potential (assumed to be zero) determines whether the
synapse is excitatory ($v_{\text{syn}} >0$) or inhibitory ($v_{\text{syn}} < 0$). 
The effect of some synapses can be described with a function that fits the shape of the
post-synaptic response due to the arrival of action potential at the
pre-synaptic release site.  A post-synaptic conductance change $g(t)$
would then be given by $g(t) = k s(t-T)$ for $t \geq T$, where $T$ is the arrival time of a pre-synaptic action potential, $s(t)$ fits the
shape of a realistic post-synaptic conductance, and $k$ is a constant. A common (normalised) choice for $s(t)$ is the $\alpha$-function:
\begin{equation}
s(t) = \alpha^2 t \e^{-\alpha t} \Theta(t) ,
\label{alpha}
\end{equation}
where $\Theta$ is a Heaviside step function.
The conductance change arising from a train of
action potentials, with firing times $T^m$, is given by
\begin{equation}
g(t) = k  \sum_m s (t-T^m) .
\label{g}
\end{equation}
If $s$ is the Green's function of a linear differential operator, so that $Q s =\delta$, then we may write (\ref{g}) in the equivalent form 
\begin{equation}
Q g = k \sum_m \delta(t-T^m) .
\label{gg}
\end{equation}
This is indeed the case for the choice (\ref{alpha}) for which 
\begin{equation}
Q=\left (1 + \frac{1}{\alpha} \FD{}{t} \right )^2 .
\label{Q}
\end{equation}

In many neural population models it is assumed that the interactions are mediated by firing rates rather than action potentials (spikes) \textit{per se}.
To see how this might arise we perform a short-time average of (\ref{gg}) over some time-scale $\tau$ and assume that $s$ is sufficiently \textit{slow} so that $\langle Q g \rangle_t$ is approximately constant, where
\begin{equation}
\langle x \rangle_t  = \frac{1}{\tau} \int_{t-\tau}^t x(t') \d t' ,
\end{equation}
then we have that $Q g = k f$, where $f$ is the instantaneous firing rate (number of spikes per time $\Delta$).  For a single neuron (real or synthetic) experiencing a constant drive it is natural to assume that this firing rate is a function of the drive alone.  If for the moment we assume that a neuron spends most of its time
close to rest such that $v_{\text{syn}}-v \approx v_{\text{syn}}$, and absorb a factor $v_{\text{syn}}$ into $k$, then for synaptically interacting neurons this drive is directly proportional to the conductance state of the presynaptic neuron.
Thus for a single population of identically and globally coupled neurons operating synchronously we are led naturally to equations like:
\begin{equation}
Q g = \kappa f(g) ,
\label{neuralmass}
\end{equation}
for some strength of coupling $\kappa$.
A common choice for the \textit{population} firing rate function is the sigmoid
\begin{equation}
f(g) = \frac{f_0}{1+\e^{-r(g-g_0)}}, 
\end{equation}
which saturates to $f_0$ for large $g$. This functional form, with threshold $g_0$ and steepness parameter $r$, is
not derived from a biophysical model, rather it is seen as a physiologically consistent choice.  The extension to multiple interacting populations is straight forward, and the popular Jansen-Rit model \cite{Jansen95}, provides a classic example of such a generalisation.  This can be written in the form
\begin{equation}
Q_{E} g_P = \kappa_P f(g_E -g_I) , \quad Q_{E} g_E = \kappa_E f(w_1 g_P) +A , \quad Q_{I}g_I = \kappa_I f(w_2 g_P) ,
\end{equation}
which describes a network of interacting pyramidal neurons (P), inhibitory interneurons (I) and excitatory interneurons (E).  Here, $Q_a$ is given by (\ref{Q}) under the replacement $\alpha \rightarrow \alpha_a$ for $a \in \{E,I\}$, $w_{1,2}$, $\kappa_{E,I,P}$ are constants, and $A$ is an external input.  It has been used to model both normal and epileptic patterns of cortical activity and its bifurcation structure has been systematically analysed in \cite{Spiegler2010,Touboul2011}.  Despite its usefulness in describing certain large scale brain rhythms, and especially alpha ($8-13$ Hz), it suffers the same deficiencies as all other neural mass models, namely it cannot track the level of synchrony \textit{within} a neuronal population.

\section{$\theta$-neuron network and reduction}
\label{sec:thetanetwork}

The $\theta$-neuron model or Ermentrout-Kopell canonical model is now widely known throughout computational neuroscience as a parsimonious model for capturing the firing and response properties of a cortical cell \cite{Ermentrout1986}.  It is described by a purely one dimensional dynamical system evolving on a circle according to
\begin{equation}
\FD{}{t}{\theta} = (1-\cos\theta) + (1+\cos\theta)\eta , \qquad \theta \in [-\pi,\pi),
\label{thetanetwork}
\end{equation}
where $\eta$ represents a constant drive.
For $\eta <0$ the $\theta$-neuron supports a pair of equilibria $\theta_\pm$, with $\theta_+<0$ and $\theta_->0$, and no equilibria for $\eta >0$.  In the former case the equilibria at $\theta_+$ is stable and the one at $\theta_-$ unstable. 
In neurophysiological terms, the unstable fixed point at $\theta_-$ is a threshold for the neuron model.  Any initial conditions with $\theta \in (\theta_+,\theta_-)$ will be attracted to the stable equilibrium, while initial data with $\theta > \theta_-$ will make a large excursion around the circle before returning to the rest state.  For $\eta >0$ the $\theta$-neuron oscillates with frequency $2\sqrt{\eta}$.  When $\eta=0$ the $\theta$-neuron is poised at a saddle-node on an invariant circle (SNIC) bifurcation.

A network of $\theta$-neurons can be described with the introduction of an index $i=1,\ldots, N$ and the replacement $\eta \rightarrow \eta_i + I_i$, where $I_i$ describes the synaptic input current to neuron $i$.  For a globally coupled network this can be written in the form $I_i = g(t) (v_{\text{syn}}-v_i)$ for some global conductance $g$ and local voltage $v_i$.  As a model for the conductance we take the form used in (\ref{gg}) and write
\begin{equation}
Q g(t) = \frac{k}{N} \sum_{j=1}^N \sum_{m \in \ZSet} \delta (t-T_j^m) ,
\label{g_diff}
\end{equation}
where $T_j^m$ is the $m$th firing time of the $j$th neuron.  These are defined to happen every time $\theta_j$ increases through $\pi$.  It is well known that the $\theta$-neuron model is formally equivalent to a quadratic integrate-and-fire model for voltage dynamics \cite{Latham2000} under the transformation $v_i = \tan(\theta_i/2)$ (so that $\cos \theta_i = (1-v_i^2)/(1+v_i^2)$ and $\sin \theta_i = 2v_i/(1+v_i^2)$).  This voltage relationship allows us to write the network dynamics as
\begin{align}
\FD{}{t}{\theta}_i &= (1-\cos\theta_i) + (1+\cos\theta_i)(\eta_i + g(t)v_{\text{syn}}) - g(t)\sin\theta_i, \label{thetanetwork} \\
Q g & = 2 \frac{k}{N} \sum_{j=1}^N P(\theta_j). 
\label{QgP}
\end{align}
Here $P(\theta) = \delta(\theta-\pi)$ and is periodically extended such that $P(\theta)=P(\theta+2 \pi)$, and we have used the result that $\delta(t-T_j^m) = \delta(\theta_j(t)- \pi) |\dot{\theta}_j(T_j^m)|$.  For the case that $v_{\text{syn}} \gg v_i$, $Q=1$, and $P(\theta)$ has a shape of the form $(1-\cos(\theta))^n$ for some positive integer $n$ we recover the model of Luke \textit{et al}. \cite{Luke2013}.  In this case these authors have shown how to obtain an exact mean-field reduction making use of the Ott-Antonsen (OA) ansatz.  The same style of reduction has also been used by Paz\'{o} and Montbri\'{o} to study pulse-coupled Winfree networks \cite{Pazo2014}.
The OA anstaz was originally used to find solutions on a reduced invariant manifold of the Kuramoto model \cite{Kuramoto91}, and essentially assumes that the distribution of phases as $N \rightarrow \infty$ has a simple unimodal shape, capable of describing synchronous (peaked) and asynchronous (flat) distributions.  In the following we show how their reduction approach extends to the more biologically realistic model described by (\ref{thetanetwork})-(\ref{QgP}) that includes synaptic reversal potentials and causal non-instantaneous synaptic  responses.  We note that even in the limit of fast synaptic interactions we do not recover models of the type described in \cite{Luke2013,Pazo2014,Montbrio2015} due to our focus on conductance changes and the inclusion of voltage shunts.

\subsection{Mean field reduction}
\label{sec:MFR}

In the following we shall choose the background drives $\eta_i$ to be random variables drawn from a Lorentzian distribution
$L(\eta)$ with
\begin{equation}
L(\eta) = \frac{1}{\pi}\frac{\Delta}{(\eta-\eta_0)^2+\Delta^2},
\label{Lorentzian}
\end{equation}
where $\eta_0$ is the centre of the distribution and $\Delta$ the width at half maximum.  For the choice of $Q$ we shall take equation (\ref{Q}).
In the coupled network, and if the frequencies of the individual neurons are similar enough, then one may expect some degree of phase locking (ranging from synchrony to asynchrony), itself controlled in part by the the time-to-peak, $1/\alpha$, of the synaptic filter.
In the limit $N \rightarrow \infty$ the state of the network at time $t$ can be described by a continuous probability distribution function 
$\rho(\eta,\theta,t)$, which satisfies the continuity equation (arising from the conservation of oscillators):
\begin{equation}
\PD{}{t} \rho +\PD{}{\theta} \rho c = 0, \qquad c = \lim_{N \rightarrow \infty} \frac{1}{N} \sum_{j=1}^N \FD{}{t}\theta_j.
\label{continuity}
\end{equation}
The global drive to the network given by the right hand side of (\ref{QgP}) can be constructed as
\begin{equation}
\lim_{N \rightarrow \infty} \frac{1}{N} \sum_{j=1}^N P(\theta_j) = \int_0^{2 \pi} \d \theta \int_{-\infty}^\infty \d \rho(\eta,\theta,t) P(\theta) .
\end{equation}
Hence,
\begin{align}
c &= (1-\cos\theta) + (1+\cos\theta)(\eta + g v_{\text{syn}}) - g\sin\theta, \label{v} \\
Q g &= \frac{k}{\pi}  \sum_{m \in \ZSet} \int_0^{2 \pi} \d \theta \int_{-\infty}^\infty \d \rho(\eta,\theta,t) \e^{im (\theta-\pi)} ,
\label{Qgpp}
\end{align}
where we have used the result that $2 \pi P(\theta) = \sum_{m \in \ZSet} \e^{im (\theta-\pi)}$.  The formula for $c$ above may be written conveniently in terms of $\e^{\pm i \theta}$ as
\begin{equation}
c =  f \e^{i\theta} + h + \overline{f} \e^{-i\theta},
\label{eq:AppendixC_simplified_velocity}
\end{equation}
where $f = ((\eta-1)+v_{\text{syn}}g+ig)/2$ and $h  = (\eta+1)+v_{\text{syn}}g$, and $\overline{f}$ denotes the complex conjugate of $f$.

The OA ansatz assumes that $\rho (\eta,\theta,t)$ has the product structure $\rho(\eta,\theta,t) = L(\eta) F(\eta,\theta,t)$.
Since $F(\eta,\theta,t)$ should be $2\pi$ periodic in $\theta$ it can be written as a Fourier series:
\begin{equation}
F(\eta,\theta,t) = \frac{1}{2 \pi} \left \{ 1 + \sum_{n=1}^\infty F_n (\eta,t) \e^{i n \theta} + \text{cc} \right \},
\end{equation}
where cc denotes complex conjugate.  The insight in \cite{Ott2008} was to restrict the Fourier coefficients such that
$F_n (\eta,t) = a(\eta,t)^n$, where $|a(\eta,t)| \leq 1$ to avoid divergence of the series.  There is also a further requirement that $a(\eta,t)$ can be analytically continued from real $\eta$ into the complex $\eta$-plane, and that this continuation has no singularities in the lower half $\eta$-plane, and that $|a(\eta,t)| \rightarrow 0$ as $\text{Im} \, \eta \rightarrow -\infty$.  If we now substitute (\ref{eq:AppendixC_simplified_velocity}) into the continuity equation (\ref{continuity}), use the OA ansatz, and balance terms in $\e^{i \theta}$ we obtain an evolution equation for $a(\eta,t)$ as
\begin{equation}
\PD{}{t} a -  i a^{2} f - ia h - i \overline{f} = 0 .
\label{alphadot}
\end{equation}
It is now convenient to introduce the Kuramoto order parameter 
\begin{equation}
Z(t) = \int_0^{2 \pi} \d \theta \int_{-\infty}^\infty \d \eta  \rho(\eta,\theta,t)  \e^{i \theta}, 
\end{equation}
where $|Z| \leq 1$.  Using the OA ansatz (and using the orthogonality properties of $\e^{i \theta}$, namely $\int_{0}^{2 \pi} \e^{i p \theta} \e^{i q \theta} \d \theta = 2 \pi \delta_{p+q,0}$) we then find that
\begin{equation}
\overline{Z}(t) = \int_{-\infty}^\infty \d \eta  L(\eta) a(\eta,t) .
\label{rcc}
\end{equation}
By noting that the Lorentzian (\ref{Lorentzian}) has simple poles at $\eta_\pm =\eta_0 \pm i \Delta$ the integral in (\ref{rcc}) may be performed by choosing a large semi-circle contour in the lower half $\eta$-plane.  This yields $\overline{Z}(t)=a(\eta_-,t)$, giving $Z(t)=a(\eta_+,t)$.
Hence, the dynamics for $g$ given by (\ref{Qgpp}) can be written as $Q g = \kappa f(Z)$, where $\kappa=k/\pi$ and 
\begin{equation}
f(Z) =  \sum_{m \in \ZSet} (-Z)^m =  \frac{1-\left|Z\right|^2}{1+Z+\overline{Z}+\left|Z\right|^2}, \qquad |Z|<1 .
\label{nextgen1}
\end{equation}
The dynamics for $Z$ is obtained from (\ref{alphadot}) as $\d Z /\d t = \mathcal{F}(Z;\eta_0,\Delta)+\mathcal{G}(Z,g;v_{\text{syn}})$, where 
\begin{align}
\mathcal{F}(Z;\eta_0,\Delta)&=   -i\frac{(Z-1)^2}{2}+\frac{(Z+1)^2}{2}\left[-\Delta + i \eta_0 \right],
\label{nextgen2} \\
\mathcal{G}(Z,g;v_{\text{syn}})&=   i\frac{(Z+1)^2}{2}  v_{\text{syn}} g  - \frac{Z^2-1}{2} g.
\label{nextgen3}
\end{align}
Here we may view (\ref{nextgen2}) as describing the intrinsic population dynamics and (\ref{nextgen3}) the dynamics induced by synaptic coupling.
Thus the form of the mean field model is precisely that of a neural mass model as given by equation (\ref{neuralmass}).  Importantly the firing rate $f$ is a derived quantity that is a real function of the complex Kuramoto order parameter for synchrony.  This in turn is described by a complex ODE with parameters from the underlying microscopic model.

\begin{figure}[htbp]
\begin{center}
\includegraphics[width=0.45\linewidth]{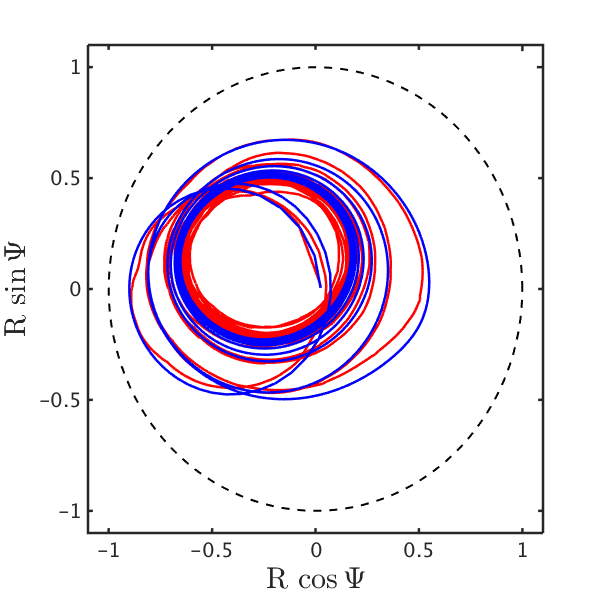} \hfill
\includegraphics[width=0.45\linewidth]{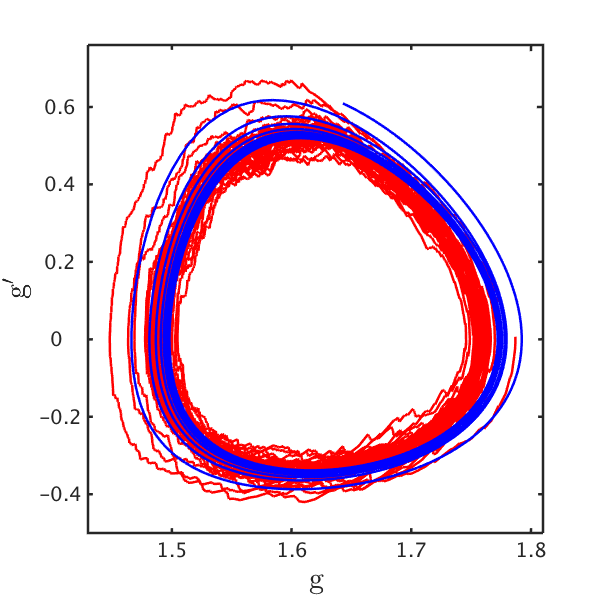}
\caption{Comparison between the reduced mean field model (blue) and a simulation of a network of $500$ $\theta$-neurons (red). 
Left:  Phase plane projection for the Kuramoto order parameter $Z=R \e^{i \Psi}$.
Right: Phase plane projection for the synaptic conductance $g$. 
Here $\eta_0=20$, $\Delta=0.5$, $v_{\text{syn}}=-10$ , $\kappa=1$, $\alpha=0.95$.
\label{Fig:MF_v_500_neurons}
}
\end{center}
\end{figure}
To illustrate the validity of the reduction presented above we show a simulation of a network with $N=500$ $\theta$-neurons and the mean field model in Fig.~ \ref{Fig:MF_v_500_neurons}.  Here we plot the real and imaginary parts of $Z$ which we write in the form $R \e^{i \Psi}$ for the mean field reduction and calculate as $Z=N^{-1} \sum_{j=1}^N \e^{i\theta_j}$ for the finite size network simulation.  It is abundantly clear that the two realisations agree very well. If the size of the population in the large scale simulations is reduced then one can more easily see finite size fluctuations as expected.

\section{Next generation neural mass model: analysis}
\label{sec:nextgeneration}

The mean field model derived in \S \ref{sec:MFR} is a natural candidate for a next generation neural mass model.  It generalises the form of the phenomenological neural mass model whilst maintaining contact with biological reality in that it preserves the notion of both population rate \textit{and} synchrony.  
An almost identical model has recently been discussed by Laing \cite{Laing2016}, although here the focus was on a first order synapse model (namely $Q=(1+\alpha^{-1} \d /\d t)$) with no provision for synaptic reversal potentials.
In mathematical terms we are now faced with understanding the dynamics of a coupled system of ODEs given by
\begin{equation}
Qg=\kappa f(Z),\qquad \FD{}{t} Z = \mathcal{F}(Z;\eta_0,\Delta) + \mathcal{G}(Z,g;v_{\text{syn}}),
\label{nextgen}
\end{equation}
with $f$, $\mathcal{F}$ and $\mathcal{G}$ given by (\ref{nextgen1}), (\ref{nextgen2}) and (\ref{nextgen3}) respectively, and $Q$ a linear differential operator such as given by (\ref{Q}).  One practical way to assess the emergent behaviour of the model (\ref{nextgen}) under parameter variation is through numerical bifurcation analysis.  We now pursue this for (\ref{nextgen}), as well as for its natural extension to cover two interacting populations.

\subsection{Bifurcation diagrams}
\label{Sec:Bif}

\begin{figure}[htbp]
\begin{center}
\includegraphics[width=0.45\linewidth]{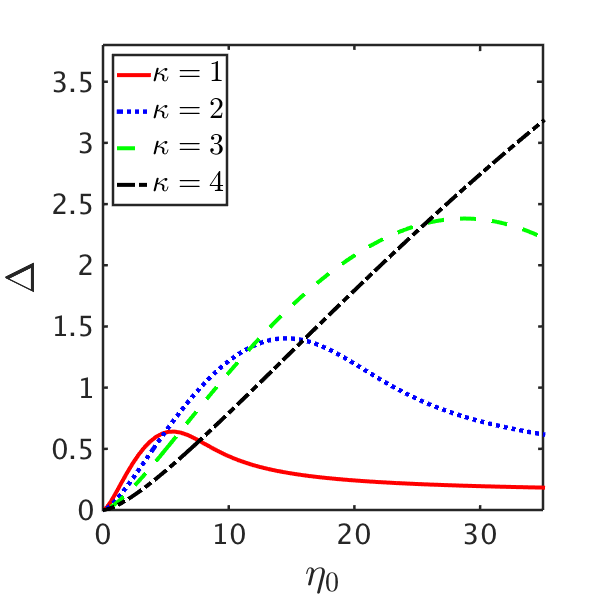} \hfill
\includegraphics[width=0.45\linewidth]{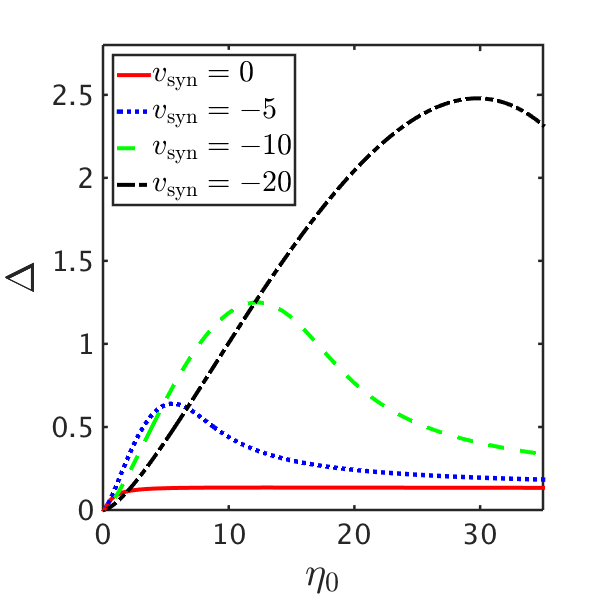}
\caption{Two parameter continuation of a Hopf bifurcation in the single population model described by (\ref{nextgen}) using (\ref{Q}) with $\alpha=1$.
Left:  Curves obtained for various values of $\kappa$ with $v_{\text{syn}}=-5$.  Right: Curves obtained under for various values of $v_{\text{syn}}$ with $\kappa=1$. 
In both diagrams the area under the curves represents the parameter window for oscillatory behaviour.
\label{Fig:OnePop}}
\end{center}
\end{figure}

We first consider the case of a purely inhibitory population.  Using {\sc XPPAUT} \cite{Ermentrout02} we find that for a wide range of system parameters it is possible to find a Hopf bifurcation of a steady state to a periodic orbit under parameter variation of $\eta_0$ (controlling the mean value of the background drive).  To illustrate the relatively large volume of parameter space that can support oscillations by this mechanism we show a two parameter continuation of the Hopf bifurcation in $(\Delta, \eta_0)$ (controlling the shape of the distribution (\ref{Lorentzian})) in Fig.~\ref{Fig:OnePop}, for several values of the coupling strength $\kappa$ and reversal potential $v_{\text{syn}}$.  Since the OA ansatz does not hold for $\Delta=0$ (so that some degree of heterogeneity must be present) and oscillations are only possible for $\eta_0>0$, the bifurcation diagrams are only presented for the scenario $\Delta, \eta_0 >0$.   

Suppose now that we have two populations, one excitatory and one inhibitory, with reciprocal connections.
Introducing the labels $E$ and $I$ for each population then the natural generalisation of (\ref{nextgen}) is
\begin{equation}
Q_{ab} g_{ab} = \kappa_{ab} f(Z_b), \qquad \FD{}{t} Z_a = \mathcal{F}_a(Z_a) + \sum_b \mathcal{G}_b(Z_a,g_{ab}),
\label{TwoPop}
\end{equation}
where $a,b \in \{E,I\}$. Here, $Q_{ab}$ is obtained from (\ref{Q}) under the replacement $\alpha \rightarrow \alpha_{ab}$, $\mathcal{F}_a(Z_a) = \mathcal{F}(Z_a;\eta_0^a,\Delta^a)$ and $\mathcal{G}_b(Z_a,g_{ab}) = \mathcal{G}(Z_a, g_{ab};v_{syn}^{ab})$.
\begin{figure}[htbp]
\begin{center}
\includegraphics[width=0.9\linewidth]{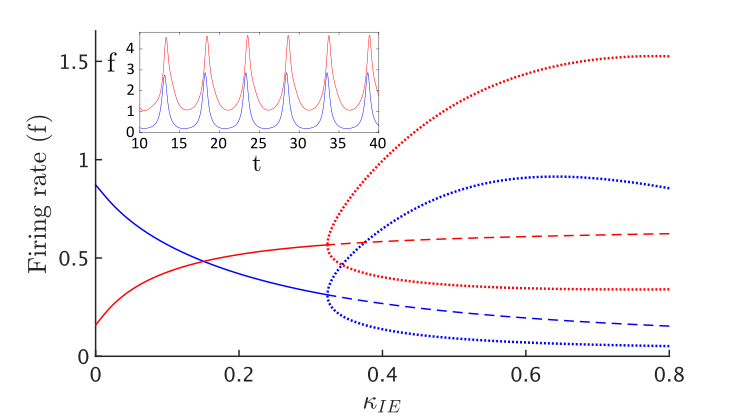}
\caption{Bifurcation diagram  for the reciprocally connected PING network defined by (\ref{TwoPop}) under variation of $\kappa_{IE}$, for both the excitatory (blue) and inhibitory (red) populations. Solid lines: stable; dashed lines: unstable. Circles show maximum and minimum values of $f(Z_E)$ and $f(Z_I)$ over one period of oscillation when no steady states are stable. The inset shows a PING rhythm for $\kappa_{IE}=0.65$. Parameters: $\alpha_{EI}=0.8$, $\alpha_{IE}=10$, $\kappa_{EI}=0.5$, $v_{\text{syn}}^{EI}=-10$, $v_{\text{syn}}^{IE}=10$, $\eta_0^E=10$, $\eta_0^I=0$, $\Delta^E=\Delta^I=0.5$, $\kappa_{EE}=\kappa_{II}=0$.
\label{Fig:TwoPopa}}
\end{center}
\end{figure}
The system of equations (\ref{TwoPop}) generalises those recently presented by Laing \cite{Laing2016} (to include reversal potentials, higher order synapses and self coupling), who highlighted the ability of such networks to produce a so-called pyramidal-interneuronal network gamma (PING) rhythm \cite{Borgers2003}, as shown in the inset of Fig.~\ref{Fig:TwoPopa}.  Here we also show a bifurcation digram as a function of $\kappa_{IE}$, when $\kappa_{EE}=\kappa_{II}=0$, which shows that periodic behaviour can be destroyed in a supercritical Hopf bifurcation as $\kappa_{IE}$ is decreased.  In Fig.~\ref{Fig:TwoPopb} we show bifurcation diagrams under the variation of $\kappa_{EI}$ and $\eta_0^I$.  
We see that when $\kappa_{EI}$ is decreased periodic behaviour can be destroyed in a supercritical Hopf bifurcation. With an increase in $\eta_0^I$ it is also possible to generate a supercritical Hopf bifurcation to terminate the PING rhythm.

\begin{figure}[htbp]
\begin{center}
\includegraphics[width=0.47\linewidth]{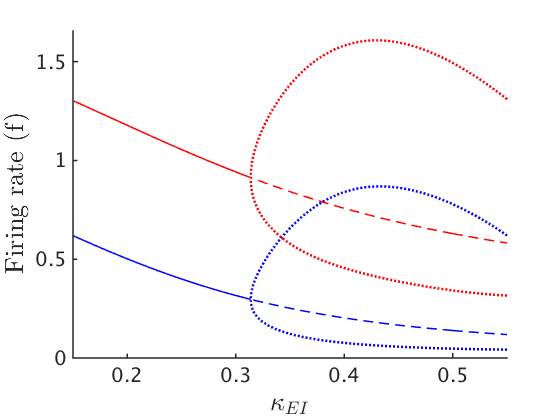} \hfill
\includegraphics[width=0.47\linewidth]{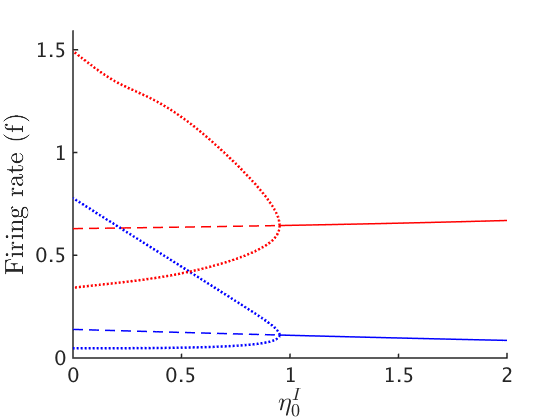}
\caption{Corresponding bifurcation diagrams to Fig.~\ref{Fig:TwoPopa} under variation in $\kappa_{EI}$ (left) and $\eta_0^I$ (right), for both the excitatory (blue) and inhibitory (red) populations.  Note that PING rhythms can be terminated by either decreasing the strength of coupling to the excitatory population from the inhibitory population or increasing the natural frequency of the inhibitory population.  Parameters as in Fig.~\ref{Fig:TwoPopa} with $\kappa_{IE} = 0.9$ for both.
\label{Fig:TwoPopb}}
\end{center}
\end{figure}

\begin{figure}[htbp]
\begin{center}
\includegraphics[width=0.70\linewidth]{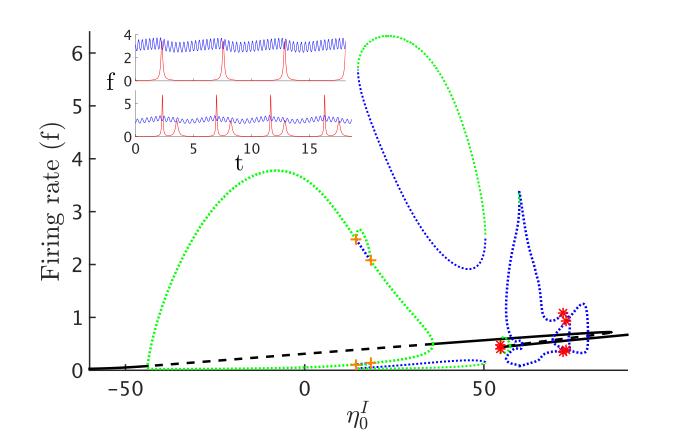}

\caption{Bifurcation diagram for $\eta_0^I$ showing the exotic behaviour of the two population model when self coupling is reintroduced. Solid lines: stable; dashed lines: unstable; green (blue) dotted line: stable (unstable) oscillations; orange crosses: period doubling bifurcations; red stars: torus bifurcations. Of particular interest is the appearance/disappearance of an isola at $\eta_0^I\simeq 15-50$. The insets shows the firing rate for the excitatory (blue) and inhibitory (red) populations.  In the upper inset $\eta_0^I=-20$ and in the lower inset $\eta_0^I=25$.  Other parameters: $\alpha_{EE}=1$, $\alpha_{EI}=0.7$, $\alpha_{IE}=1.4$, $\alpha_{II}=0.4$, $\kappa_{EE}=1.5$, $\kappa_{EI}=2$, $\kappa_{IE}=1$, $\kappa_{II}=3$, $v_{\text{syn}}^{EE}=10$, $v_{\text{syn}}^{IE}=-v_{\text{syn}}^{EI}=8$, $v_{\text{syn}}^{II}=-12$, $\eta_0^E=20$, $\Delta^E=\Delta^I=0.5$.
\label{Fig:TwoPopc}}
\end{center}
\end{figure}

The inclusion of self coupling leads to a wide variety of bifurcations, as seen in Fig. \ref{Fig:TwoPopc}. As $\eta_0^I$ is increased we observe the appearance of oscillatory behaviour through a supercritical Hopf bifurcation, which is destroyed at a second supercritical Hopf bifurcation when $\eta_0^I$ is increased further. Note the appearance/disappearance of period doubling through two period doubling bifurcations on this branch of periodic orbits. We also observe the appearance and disappearance of an isola of periodic orbits through two saddle node bifurcations of periodic orbits. The first saddle node occurs before the second Hopf bifurcation, i.e. there exists two stable periodic orbits for this window of parameter space. Further increasing $\eta_0^I$ leads to another  saddle node bifurcation of periodic orbits, shortly followed by a torus bifurcation and then a saddle-node on invariant circle bifurcation which destroys the unstable branch of the periodic orbit. The stable branch of the periodic orbit is destroyed at a supercritical Hopf bifurcation, as $\eta_0^I$ is increased further. Along the unstable fixed point branch there are four Hopf bifurcations all of which either create or destroy unstable periodic behaviour. Between the second and third of these bifurcations there are two torus bifurcations, one on each periodic orbit.

The inset in Fig. \ref{Fig:TwoPopc} shows the behaviour of the system for $\eta_0^I=-20$ and $\eta_0^I=25$ respectively. In both cases the excitatory population has two frequencies and the lower frequency is synchronised to the inhibitory population. For  $\eta_0^I=25$, the system follows the orbit created by the isola, and there are two peaks in the (inhibitory) firing rate per period. 

\subsection{Event related synchronisation and desynchronisation}

\begin{figure}[htbp]
\begin{center}
\includegraphics[width=0.45\linewidth]{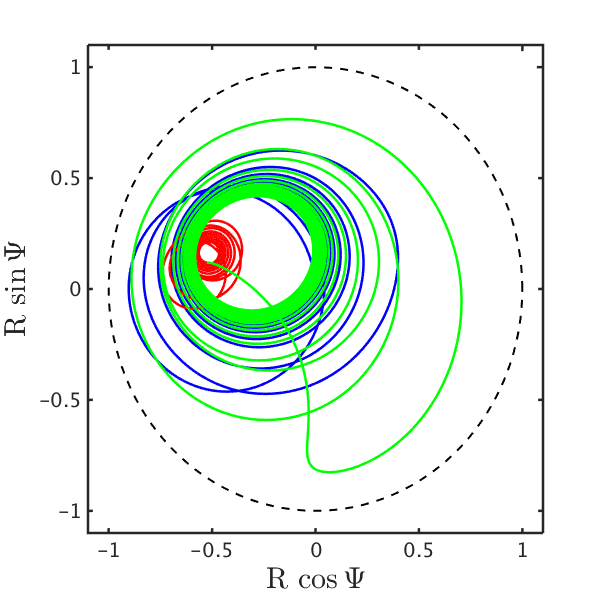} \hfill
\includegraphics[width=0.45\linewidth]{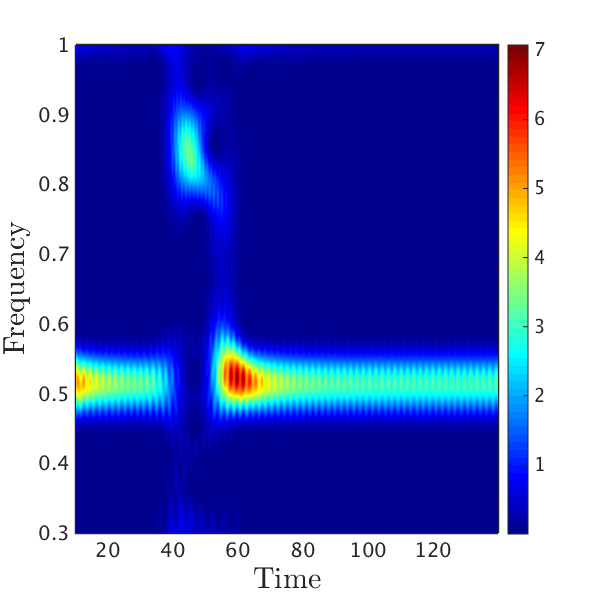}
\caption{
Left: Phase plane for $Z=R \e^{i \Psi}$, demonstrating the behaviour of the system (\ref{nextgen}) in response to a drive in the form of a smoothed rectangular pulse. The blue curve represents the system before the pulse arrives, as it settles to its non-perturbed dynamics ($t<T$),the red curve demonstrates how the system behaves when the pulse is switched on (at $t=T$) and the green how the system reacts once the drive is switched off (at $t=T+\tau$). 
Right:  The corresponding spectrogram of the synaptic current demonstrating \textit{rebound} (an enhanced spectral peak on cessation of the applied pulse).
Parameter values as in Fig.~\ref{Fig:MF_v_500_neurons}, with $T=40$, $\tau=12$, $\eta_0=21.5$, $\alpha_D = 6$ and $\sigma = 15$. 
\label{Fig:Drive}}
\end{center}
\end{figure}

Here we show that a time varying input to a single population model given by (\ref{nextgen}) can both disrupt and enhance the degree of synchrony within the population.  We include such a drive under the replacement $\eta_0 \rightarrow \eta_0 + J(t)$ (describing a homogeneous drive to the microscopic system), where $J(t)$ is a smoothed rectangular pulse of the form $J(t) = \int_{-\infty}^ t \eta_D(t-s) A(s) \d s$, for $A(t) = \sigma \Theta(t-T)\Theta(T+\tau-t)$ and $\eta_D(t)$ is an $\alpha$-function of the form (\ref{alpha}) under the replacement $\alpha \rightarrow \alpha_D$. 
In Fig.~\ref{Fig:Drive} we show that for the inhibitory population considered in \S \ref{sec:MFR} operating in its oscillatory regime that such a drive can initially cause a population to desynchronise (during the pulse) though that upon termination of the pulse the system can \textit{rebound} and generate a stronger spectral power peak than seen before the presentation of the pulse (before relaxing back to the undriven periodic orbit).
Thus, in contrast to a standard neural mass model, the mean field model (\ref{nextgen}) is mechanistically able to support the phenomena of ERD and ERS.  A more thorough discussion of this observation can be found in Byrne \textit{et al}. \cite{Byrne2016}, where the model is used to describe magnetoencephalography (MEG) data for movement related beta decrease and post-movement beta rebound \cite{Pfurtscheller1999}.

\section{Discussion}
\label{sec:discussion}
The desire to understand large scale brain dynamics as observed using EEG, MEG and fMRI has prompted the increasing use of computational models \cite{Bojak2014}.  Many of these approaches, such as exemplified by the Virtual Brain project \cite{Sanz-Leon2015}, make use of networks of interconnected neural mass models.  However, the inability of a single neural mass model to support the well documented phenomena of ERS and ERD reminds us that these phenomenological models could be improved upon.  Of course, building more detailed biophysically realistic models of neurons and their interactions would improve this state of affairs, though at a price.  This being not only computational complexity but our ability to interpret the behaviour of very high dimensional models in a meaningful way.  The model that we have presented here is very much in the original spirit of neural mass modelling, yet importantly it can be interpreted directly in terms of an underlying spiking model.  Moreover the \textit{derived} structure of the macroscopic equations can be viewed as a modification of the standard neural mass framework 
whereby the firing rate of the system is now coupled to the degree of synchrony within a population.  Given the recent success of this model in explaining beta rebound \cite{Byrne2016}, we advocate strongly for its subsequent use in future population-level modelling approaches for understanding \textit{in vivo} brain activity states.  Indeed we would like to think that this is a first example of a next generation neural mass model and that there will be others to follow.  For phase oscillator single neuron models, this is intrinsically linked to the mathematical challenge of generalising the OA approach, whilst for more general conductance based neurons one might appeal to the McKean-Vlasov-Fokker-Planck approach of Baladron \textit{et al}. \cite{Baladron12}.  However, even before extending the model we have presented here to include more biological features (such as action potential generation, dendritic processing, and stochasticity) there is still much to be done in understanding the response of the model to input.  Of particular utility would be an understanding of the response to periodic forcing, as this would be a precursor to understanding patterns of phase-locking, clustering, chaos, and the multiplicity of attractors expected at the network level.  Moreover, given that neural mass models are themselves the building blocks of neural field models \cite{Coombes14} it would be interesting to pursue the analysis of bumps, waves and patterns in continuum versions of the model presented here, along the lines recently developed by Laing for both synaptic and gap junction coupled systems \cite{Laing2015,Laing2016}.

\begin{acknowledgement}
SC was supported by the European Commission through the FP7 Marie Curie Initial Training Network 289146, NETT: Neural Engineering Transformative Technologies.
\end{acknowledgement}

\bibliographystyle{unsrt}
\bibliography{MeanField}

\begin{thebibliography}{10}

\bibitem{Wilson72}
H~R Wilson and J~D Cowan.
\newblock Excitatory and inhibitory interactions in localized populations of
  model neurons.
\newblock {\em Biophysical Journal}, 12:1--24, 1972.

\bibitem{Zetterberg1978}
L~H Zetterberg, L~Kristiansson, and K~Mossberg.
\newblock Performance of a model for a local neuron population.
\newblock {\em Biological Cybernetics}, 31:15--26, 1978.

\bibitem{daSilva1974}
F~H~Lopes da~Silva, A~Hoeks, H~Smits, and L~H Zetterberg.
\newblock Model of brain rhythmic activity: {T}he alpha-rhythm of the thalamus.
\newblock {\em Kybernetik}, 15:27--37, 1974.

\bibitem{daSilva1976}
F~H~Lopes da~Silva, A~van Rotterdam, P~Barts, E~van Heusden, and W~Burr.
\newblock Models of neuronal populations: the basic mechanisms of rhythmicity.
\newblock {\em Progress in Brain Research}, 45:281--308, 1976.

\bibitem{Jansen95}
B~H Jansen and V~G Rit.
\newblock Electroencephalogram and visual evoked potential generation in a
  mathematical model of coupled cortical columns.
\newblock {\em Biological Cybernetics}, 73:357--366, 1995.

\bibitem{Wendling2015}
F~Wendling, P~Benquet, F~Bartolomei, and V~Jirsa.
\newblock Computational models of epileptiform activity.
\newblock {\em Journal of Neuroscience Methods}, 260:233--251, 2016.

\bibitem{Dafilis09}
M~P {Dafilis}, F~{Frascoli}, P~J {Cadusch}, and D~T~J {Liley}.
\newblock {Chaos and generalised multistability in a mesoscopic model of the
  electroencephalogram}.
\newblock {\em Physica D}, 238:1056--1060, 2009.

\bibitem{Freeman1992}
W~J Freeman.
\newblock Tutorial on neurobiology: From single neurons to brain chaos.
\newblock {\em International Journal of Bifurcation and Chaos}, 2:451--482,
  1992.

\bibitem{Sotero2007}
R~C Sotero, N~J Trujillo-Barreto, Y~Iturria-Medina, F~Carbonell, and J~C
  Jimenez.
\newblock Realistically coupled neural mass models can generate {EEG} rhythms.
\newblock {\em Neural Computation}, 19:478--512, 2007.

\bibitem{Spiegler2011}
A~Spiegler, T~R Kn\"{o}sche, K~Schwab, J~Haueisen, and F~M Atay.
\newblock Modeling brain resonance phenomena using a neural mass model.
\newblock {\em PLoS Computational Biology}, 7(12):e1002298, 2011.

\bibitem{Deco11}
G~Deco, V~K Jirsa, and A~R McIntosh.
\newblock Emerging concepts for the dynamical organization of resting-state
  activity in the brain.
\newblock {\em Nature Reviews Neuroscience}, 12:43--56, 2011.

\bibitem{Valdes-Sosa2009}
P~Valdes-Sosa, J~M Sanchez-Bornot, R~C Sotero, Y~Iturria-Medina,
  Y~Aleman-Gomez, J~Bosch-Bayard, F~Carbonell, and T~Ozaki.
\newblock {Model driven EEG/fMRI fusion of brain oscillations}.
\newblock {\em Human Brain Mapping}, 30:2701--21, 2009.

\bibitem{Moran2013}
R~Moran, D~A Pinotsis, and K~Friston.
\newblock Neural masses and fields in dynamic causal modeling.
\newblock {\em Frontiers in Computational Neuroscience}, 7(57):1--12, 2013.

\bibitem{Sanz-Leon2015}
P~Sanz-Leon, S~A Knock, A~Spiegler, and V~K Jirsa.
\newblock Mathematical framework for large-scale brain network modeling in {The
  Virtual Brain}.
\newblock {\em NeuroImage}, 111:385--430, 2015.

\bibitem{Bhattacharya2015}
B~S Bhattacharya and F~N Chowdhury, editors.
\newblock {\em Validating Neuro-Computational Models of Neurological and
  Psychiatric Disorders}.
\newblock Springer, 2015.

\bibitem{Pfurtscheller1999}
G~Pfurtscheller and F~H~Lopes da~Silva.
\newblock Event-related {EEG/MEG} synchronization and desynchronization: basic
  principles.
\newblock {\em Clinical Neurophysiology}, 110:1842--1857, 1999.

\bibitem{Ashwin2016}
P~Ashwin, S~Coombes, and R~Nicks.
\newblock Mathematical frameworks for oscillatory network dynamics in
  neuroscience.
\newblock {\em Journal of Mathematical Neuroscience}, 6(2), 2016.

\bibitem{Luke2013}
T~B Luke, E~Barreto, and P~So.
\newblock Complete classification of the macroscopic behaviour of a
  heterogeneous network of theta neurons.
\newblock {\em Neural Computation}, 25:3207--3234, 2013.

\bibitem{Montbrio2015}
E~Montbri\'o, D~Paz\'o, and A~Roxin.
\newblock Macroscopic description for networks of spiking neurons.
\newblock {\em Physical Review X}, 5:021028, 2015.

\bibitem{Spiegler2010}
A~Spiegler, S~J Kiebel, F~M Atay, and T~R Kn\"osche.
\newblock Bifurcation analysis of neural mass models: Impact of extrinsic
  inputs and dendritic time constants.
\newblock {\em NeuroImage}, 52:1041--1058, 2010.

\bibitem{Touboul2011}
J~Touboul, F~Wendling, P~Chauvel, and O~Faugeras.
\newblock Neural mass activity, bifurcations, and epilepsy.
\newblock {\em Neural Computation}, 23:3232--3286, 2011.

\bibitem{Ermentrout1986}
G~B Ermentrout and N~Kopell.
\newblock Parabolic bursting in an excitable system coupled with a slow
  oscillation.
\newblock {\em SIAM Journal on Applied Mathematics}, 46:233--253, 1986.

\bibitem{Latham2000}
P~E Latham, B~J Richmond, P~G Nelson, and S~Nirenberg.
\newblock {Intrinsic Dynamics in Neuronal Networks . I . Theory}.
\newblock {\em Journal of Neurophysiology}, 83:808--827, 2000.

\bibitem{Pazo2014}
D~Paz\'{o} and E~Montbri\'{o}.
\newblock Low-dimensional dynamics of populations of pulse-coupled oscillators.
\newblock {\em Physical Review X}, 4:011009, 2014.

\bibitem{Kuramoto91}
Y~Kuramoto.
\newblock Collective synchronization of pulse-coupled oscillators and excitable
  units.
\newblock {\em Physica D}, 50:15--30, 1991.

\bibitem{Ott2008}
E~Ott and T~M Antonsen.
\newblock Low dimensional behavior of large systems of globally coupled
  oscillators.
\newblock {\em Chaos}, 18:037113, 2008.

\bibitem{Laing2016}
C~R Laing.
\newblock {\em Computational Models of Brain and Behavior}, chapter Phase
  oscillator network models of brain dynamics.
\newblock Wiley-Blackwell, 2016.

\bibitem{Ermentrout02}
G~B Ermentrout.
\newblock {\em Simulating, analyzing, and animating dynamical systems: {A}
  guide to {XPPAUT} for researchers and students}.
\newblock {SIAM} Books, Philadelphia, 2002.

\bibitem{Borgers2003}
C~B\"orgers and N~Kopell.
\newblock Synchronization in networks of excitatory and inhibitory neurons with
  sparse, random connectivity.
\newblock {\em Neural Computation}, 15:509--538, 2003.

\bibitem{Byrne2016}
\'A Byrne, M~J Brookes, and S~Coombes.
\newblock A mean field model for movement induced changes in the $\beta$
  rhythm.
\newblock {\em NeuroImage}, submitted, 2016.

\bibitem{Bojak2014}
I~Bojak and M~Breakspear.
\newblock Neuroimaging, neural population models for.
\newblock In {\em Encyclopedia of Computational Neuroscience}, pages 1--29.
  Springer, 2014.

\bibitem{Baladron12}
J~Baladron, D~Fasoli, O~Faugeras, and J~Touboul.
\newblock Mean field description of and propagation of chaos in networks of
  {H}odgkin-{H}uxley and {F}itz{H}ugh-{N}agumo neurons.
\newblock {\em Journal of Mathematical Neuroscience}, 2(10), 2012.

\bibitem{Coombes14}
S~Coombes, P~beim Graben, R~Potthast, and J~J Wright, editors.
\newblock {\em Neural Field Theory}.
\newblock Springer Verlag, 2014.

\bibitem{Laing2015}
C~R Laing.
\newblock Exact neural fields incorporating gap junctions.
\newblock {\em SIAM Journal on Applied Dynamical Systems}, 14:1899--1929, 2015.

\end{thebibliography}

\end{document}